\documentstyle[preprint,prb,aps,fleqn]{revtex}

\begin{document}
\title{Coherent states, displaced number states and Laguerre polynomial states for
su(1,1) Lie algebra}
\author{Xiao-Guang Wang\thanks{%
email:xyw@aphy.iphy.ac.cn}}
\address{Laboratory of Optical Physics, Institute of Physics, \\
Chinese Academy of Sciences, Beijing 100080, People's Republic of China}
\date{\today}
\maketitle

\begin{abstract}
The ladder operator formalism of a general quantum state for su(1,1) Lie
algebra is obtained. The state bears the generally deformed oscillator
algebraic structure. It is found that the Perelomov's coherent state is a
su(1,1) nonlinear coherent state. The expansion and the exponential form of
the nonlinear coherent state are given. We obtain the matrix elements of the
su(1,1) displacement operator in terms of the hypergeometric functions and
the expansions of the displaced number states and Laguerre polynomial states
are followed. Finally some interesting su(1,1) optical systems are discussed.
\end{abstract}


\section{Introduction}

The harmonic oscillator is a fundamental exactly solvable physical system
and the coherent state(CS)\cite{CS} defined in this system is well studied.
The generalization of the CS  to multi-photon case\cite{MMCS} and the
extention to various systems have been made.

As for su(1,1) Lie
algebra, the Perelomov's coherent state(PCS) is well known\cite{PCS} .
The
su(1,1) Lie algebra is of great interest in quantum optics because it can
characterize many kinds of quantum optical systems. In particular, the
bosonic realization of su(1,1) describes the degenerate and non-degenerate
parametric amplifiers.

The generators of su(1,1) Lie algebra, $K_0$ and $K_{\pm },$ satisfy the
commutation relations

\begin{equation}
\lbrack K_{+},K_{-}]=-2K_0,\text{ }[K_0,K_{\pm }]=\pm K_{\pm }.
\end{equation}
Its discrete representation is

\begin{eqnarray}
K_{+}|n,k\rangle &=&\sqrt{(n+1)(2k+n)}|n+1,k\rangle , \\
K_{-}|n,k\rangle &=&\sqrt{n(2k+n-1)}|n-1,k\rangle ,  \nonumber \\
K_0|n,k\rangle &=&(n+k)|n,k\rangle .  \nonumber
\end{eqnarray}
Here $|n,k\rangle (n=0,1,2,...)$ is the complete orthonormal basis and $%
k=1/2,1,3/2,2,...$ is the Bargmann index labeling the irreducible
representation[$k(k-1)$ is the value of Casimir operator]. We
introduce the number operator ${\cal N}$ by

\begin{equation}
{\cal N}=K_0-k,{\cal N}|n,k\rangle =n|n,k\rangle .
\end{equation}

The PCS is defined as

\begin{eqnarray}
|\alpha ,k\rangle _P &=&S(\xi )|0,k\rangle \\
&=&(1-|\alpha |^2)^k\sum_{n=0}^\infty \sqrt{\frac{\Gamma (2k+n)}{\Gamma
(2k)n!}}\alpha ^n|n,k\rangle ,  \nonumber
\end{eqnarray}
where $\xi =r\exp (i\theta )$ ,$\alpha =\exp (i\theta )\tanh r$ , $\Gamma
(x) $ is the gamma function, $S(\xi )=\exp (\xi K_{+}-\xi ^{*}K_{-})$ is the
su(1,1) displacement operator$.$ There is another coherent state of su(1,1)
which is known as the Barut-Girardello(BG) coherent state(BGCS)\cite{BGCS}.
The BGCS is defined as the eigenstate of the lowering operator $K_{-}$

\begin{equation}
K_{-}|\alpha ,k\rangle _{BG}=\alpha |\alpha ,k\rangle _{BG},
\end{equation}
and it can be expressed as\cite{BGCS}

\begin{equation}
|\alpha ,k\rangle _{BG}=\sqrt{\frac{|\alpha |^{2k-1}}{I_{2k-1}(2|\alpha |)}}%
\sum_{n=0}^\infty \frac{\alpha ^n}{\sqrt{n!\Gamma (n+2k)}}|n,k\rangle ,
\end{equation}
where $I_\nu (x)$ is the first kind modified Bessel function.

The PCS is defined as the displacement operator formalism, while the BGCS as
the ladder operator formalism. We ask if the PCS admits the ladder operator
formalism? The answer is affirmative. We will discuss it in the next
section. We also give the ladder operator formalism of a general su(1,1)
quantum state and find that the PCS is a su(1,1) nonlinear coherent
state(NLCS). The complete expansion and exponential form of the su(1,1) NLCS
are obtained.

There are three definitions of coherent states, that is, (1) the
displacement operator acting on the vacuum states, (2) the eigenstates of
the annihilation operator, (3) the minimum uncertainty states. These three
definitions are identical only for the simplest harmonic oscillator. For
su(1,1) system, the PCS is defined according to the first and the BGCS to
the second. The
minimum uncertainty states(MUSs) for su(1,1) are defined as\cite{MUS}
\begin{equation}
(\mu K_{+}+\nu K_{-})|\alpha ,k\rangle _{MUS}=\alpha |\alpha ,k\rangle
_{MUS},
\end{equation}
where $\mu $ and $\nu $ are complex constants satisfying $|\mu /\nu |<1.$
One type of the MUS is the Laguerre polynomial state(LPS)\cite{LPS}. The
LPS is only given formally in the literature. We will give the expansion of
the LPS in terms of states $|n,k\rangle $ in section III. The PCS, BGCS and
MUS cover the three definations of the coherent states. In section IV, we
consider several interesting su(1,1) optical systems, namely, the
density-dependent Holstein-Promakoff system\cite{HP}, amplitude-squared
system\cite{ONE}, and two-mode system\cite{TWO}. A conclusion is given in
Sec.V.

\section{su(1,1) coherent states and nonlinear coherent states}

We consider a general state

\begin{equation}
|x,k\rangle _G=\sum_{n=0}^\infty C(n,x,k)|n,k\rangle ,
\end{equation}
where $x$ denote parameter and all the coefficients $C(n,x,k)$ are 
non-zero. Now we try to give the ladder operator formalism of the above
general state. The key point is to let the number operator ${\cal N}$ and
the operator $f({\cal N})K_{+}$ act on Eq.(8), respectively. Here $f({\cal N}%
)$ is a real function of ${\cal N}.$ The operations lead to

\begin{eqnarray}
{\cal N}|x,k\rangle _G &=&\sum_{n=1}^\infty C(n,x,k)n|n,k\rangle , \\
f({\cal N})K_{+}|x,k\rangle _G &=&\sum_{n=1}^\infty f(n)C(n-1,x,k)\sqrt{%
n(n+2k-1)}|n,k\rangle .  \nonumber
\end{eqnarray}
If we choose

\begin{equation}
f({\cal N})=\frac{C({\cal N},k,x)\sqrt{{\cal N}}}{C({\cal N}-1,k,x)\sqrt{%
{\cal N}+2k-1}},
\end{equation}
the following equation is obtained

\begin{equation}
\lbrack {\cal N}-f({\cal N})K_{+}]|x,k\rangle _G=0.
\end{equation}
This is the ladder operator formalism of the general state $|x,k\rangle _G.$

Let us examine the algebraic structure involved in the general state. Define 
${\cal A}$ as an associate algebra with generators 
\begin{equation}
{\cal N},A_{+}=\frac{C({\cal N},k,x)\sqrt{{\cal N}}}{C({\cal N}-1,k,x)\sqrt{%
{\cal N}+2k-1}}K_{+},A_{-}=(A_{+})^{\dagger }.
\end{equation}
Then it is easy to verify that these operators satisfy the following
relations 
\begin{equation}
\lbrack {\cal N},A_{\pm }]=\pm A_{\pm },\text{ }A_{+}A_{-}=S({\cal N}),\text{
}A_{-}A_{+}=S({\cal N}+1),
\end{equation}
where the function 
\begin{equation}
S({\cal N})=\frac{{\cal N}^2C^2({\cal N},k,x)}{C^2({\cal N}-1,k,x)}
\end{equation}
This algebra ${\cal A}$ is nothing but the generally deformed oscillator(GDO)%
\cite{GDO} algebra with the structure function $S({\cal N})$. So we see that
the general state $|x,k\rangle _G$ bears generally deformed oscillator
algebraic structure.

By acting the annihilation operator $K_{-}$ on Eq.(11) from left, we get

\begin{equation}
\lbrack f({\cal N}+1)({\cal N}+2k)-K_{-}]|x,k\rangle _G=0.
\end{equation}
In the derivation of the above equation, we have used the fact that the
operator $%
{\cal N}+1$ is non-zero in the whole space. The function $f({\cal N})$ is
completely determined by the coefficients of the state $|x,k\rangle _G.$
From the coefficients of the BGCS(Eq.(6)), the operator-valued function $f(%
{\cal N})=\alpha /({\cal N}+2k-1).$ It is easily seen that Eq.(15)
reduces to Eq.(5) as we expected. From the coefficients of the PCS(Eq.(4)),
we obtain the corresponding operator-valued function $f({\cal N})=\alpha .$
This simple result leads to the ladder operator formalism of the PCS

\begin{equation}
\frac 1{{\cal N}+2k}K_{-}|\alpha ,k\rangle _P=\alpha |\alpha ,k\rangle _P
\end{equation}

In fact, by direct verification, we have 
\begin{equation}
\lbrack \frac 1{{\cal N}+2k}K_{-},K_{+}]=1.
\end{equation}
Therefore, the exponential formalism of the PCS can be given by

\begin{equation}
|\alpha ,k\rangle _P=\exp (\alpha K_{+})|0,k\rangle
\end{equation}
up to a normalization constant.

Reminding the definition of nonlinear coherent states in Fock space\cite
{NLCS}, we can call the state $|\alpha ,k\rangle _P$ as a su(1,1) NLCS. The
NLCS is defined as

\begin{equation}
G({\cal N})K_{-}|\alpha ,k\rangle _{NL}=\alpha |\alpha ,k\rangle _{NL},
\end{equation}
where $G({\cal N})$ is a real function of ${\cal N}.$ The PCS and the BGCS
are recovered for the special choices of $G({\cal N})=1/({\cal N}+2k)$ and $%
G({\cal N})=1,$ respectively. Thus the two coherent states, PCS and BGCS,
are unified within the framework of the su(1,1) NLCS.

Assuming the expansion of the NLCS is

\begin{equation}
|\alpha ,k\rangle _{NL}=\sum_{n=0}^\infty D(n,\alpha ,k)|n,k\rangle
\end{equation}
and substituting it into Eq.(19), we get the following recursion relation

\begin{equation}
\frac{D(n+1,\alpha ,k)}{D(n,\alpha ,k)}=\frac \alpha {G(n)\sqrt{(n+1)(2k+n)}}%
.
\end{equation}
Eq.(21) leads to

\begin{equation}
D(n,\alpha ,k)=\frac{\alpha ^nD(0,\alpha ,k)}{G(n-1)G(n-2)...G(0)\sqrt{%
n!\Gamma (2k+n)/\Gamma (2k)}}.
\end{equation}
The combination of Eq.(20) and (22) gives the expansion of the su(1,1) NLCS,

\begin{eqnarray}
|\alpha ,k\rangle _{NL} &=&D(0,\alpha ,k)\sum_{n=0}^\infty \frac{\alpha
^nD(0,\alpha ,k)\sqrt{\Gamma (2k)}}{G(n-1)G(n-2)...G(0)\sqrt{n!\Gamma (2k+n)}%
}|n,k\rangle , \\
&=&D(0,\alpha ,k)\sum_{n=0}^\infty \frac{\alpha ^n\Gamma (2k)K_{+}^n}{%
G(n-1)G(n-2)...G(0)n!\Gamma (2k+n)}|0,k\rangle .  \nonumber
\end{eqnarray}

The coefficient $D(0,\alpha ,k)$ can be determined by normalization. Let $G(%
{\cal N})=1/({\cal N}+2k),$ we naturally reduce Eq.(23) to Eq.(4) up to a
normalization constant.

One can show that

\begin{eqnarray}
{\cal N}K_{+} &=&K_{+}({\cal N}+1),f({\cal N})K_{+}=K_{+}f({\cal N}+1), \\
\lbrack f({\cal N})K_{+}]^n &=&(K_{+})^nf({\cal N}+1)f({\cal N}+2)...f({\cal %
N}+n).  \nonumber
\end{eqnarray}
Then as a key step, by using Eq.(24) with

\begin{equation}
f({\cal N})=\frac \alpha {G({\cal N}-1)({\cal N}+2k-1)},
\end{equation}
the NLCS is finally written in the exponential form

\begin{eqnarray}
|\alpha ,k\rangle _{NL} &=&D(0,\alpha ,k)\sum_{n=0}^\infty \frac 1{n!}[\frac %
\alpha {G({\cal N}-1)({\cal N}+2k-1)}K_{+}]^n|0,k\rangle ,  \nonumber \\
&=&D(0,\alpha ,k)\exp [\frac \alpha {G({\cal N}-1)({\cal N}+2k-1)}%
K_{+}]|0,k\rangle .
\end{eqnarray}

From the above equation, the exponential form of the BGCS is easily obtained
by setting $G({\cal N})=1$,

\begin{equation}
|\alpha ,k\rangle _{BG}=[\exp \frac \alpha {({\cal N}+2k-1)}K_{+}]|0,k\rangle
\end{equation}
up to a normalization constant. Let $G({\cal N})=1/({\cal N}+2k)$ in Eq.(26)$%
,$ Eq.(18) is recovered as we expected.

Actually we have

\begin{equation}
\lbrack G({\cal N})K_{-},\frac 1{G({\cal N}-1)({\cal N}+2k-1)}K_{+}]=1.
\end{equation}
By this observation, Eq.(26) is naturally obtained.

\section{Displaced number states and Laguerre polynomial states}

As a generalization of the PCS, we define the displaced number state(DNS)
for su(1,1) Lie algebra in analogous with the definition of the displaced
number state in Fock space,

\begin{equation}
|\xi ,m,k\rangle _{DN}=S(\xi )|m,k\rangle =\sum_{n=0}^\infty \langle
n,k|S(\xi )|m,k\rangle |n,k\rangle ,\text{ }\xi =r\exp (i\theta ).
\end{equation}
All the work left is to calculate the matrix elements $S_{nm}^k(\xi
)=\langle n,k|S(\xi )|m,k\rangle .$ Using the decomposed form of the
displacement operator

\begin{equation}
S(\xi )=\exp (\alpha K_{+})(1-|\alpha |^2)^{K_0}\exp (-\alpha ^{*}K_{-})%
\text{ },\alpha =\exp (i\theta )\tanh r
\end{equation}
and the relation

\begin{equation}
\exp (-\eta ^{*}K_{-})|m,k\rangle =\sum_{q=0}^m\frac{(-\eta ^{*})^{m-q}}{%
(m-q)!}\sqrt{\frac{m!\Gamma (2k+m)}{q!\Gamma (2k+q)}}|q,k\rangle
\end{equation}
we obtain the matrix elements as

\begin{eqnarray}
S_{nm}^k(\xi ) &=&(1-|\alpha |^2)^k\alpha ^n(-\alpha ^{*})^m\sqrt{m!n!\Gamma
(2k+m)\Gamma (2k+n)} \\
&&\sum_{q=0}^{\min (m,n)}\frac{(1-1/|\alpha |^2)^q}{q!(n-q)!(m-q)!\Gamma
(2k+q)}.  \nonumber
\end{eqnarray}
Using the relations

\begin{equation}
(-m)_q=(-1)^q\frac{m!}{(m-q)!},(-n)^q=(-1)^q\frac{n!}{(n-q)!},(2k)_q=\frac{%
\Gamma (2k+q)}{\Gamma (2k)},
\end{equation}
we can write the matrix elements in terms of hypergeometric function as

\begin{eqnarray}
S_{nm}^k(\xi ) &=&(1-|\alpha |^2)^k\alpha ^n(-\alpha ^{*})^m \\
&&\sqrt{\frac{\Gamma (2k+m)\Gamma (2k+n)}{\Gamma (2k)\Gamma (2k)m!n!}}\text{ 
}_2F_1(-m,-n;2k;1-\frac 1{|\alpha |^2}).  \nonumber
\end{eqnarray}
Here the hypergeometric function

\begin{equation}
_2F_1(\alpha ,\beta ;\gamma ;z)=\sum_{n=0}^\infty \frac{(\alpha )_n(\beta )_n%
}{n!(\gamma )_n}z^n,
\end{equation}
and

\begin{equation}
(x)_n=x(x+1)...(x+n-1),(x)_0\equiv 1.
\end{equation}

The combination of Eqs.(29) and (34) gives the expansion of the DNS in
terms
of the basis state $|n,k\rangle .$ It is easily checked that Eq.(29) reduces
to Eq.(4) when $m=0.$ The matrix elements abtained here are useful in the
study of su(1,1) quantum states. 

As one type of MUS for su(1,1) Lie algebra, the LPS is given by\cite{LPS},

\begin{equation}
|\alpha ,k\rangle _{LP}=C_0S(\beta )L_M(\xi \frac{K_0-k}{K_0+k-1}%
K_{+})|0,k\rangle .
\end{equation}
Here $\beta =r\exp (i\theta )$ is determined by the equation $\exp (2i\theta
)\tanh ^2r=-\nu /\mu .$ $\xi =-\exp (i\theta )\tanh (2r),$ $C_0$ can be
determined by normalization, and

\begin{equation}
L_M(x)=\sum_{n=0}^M\frac 1{n!}{%
{M \choose M-n}%
}(-1)^nx^n
\end{equation}
is the Laguerre polynomial. Using Eq.(38), we obtain the expansion of the
LPS as

\begin{eqnarray}
|\alpha ,k\rangle _{LP} &=&C_0S(\beta )\sum_{m=0}^M(-\xi )^n\frac{M!}{(M-m)!%
\sqrt{m!\Gamma (2k+m)/\Gamma (2k)}}|m,k\rangle , \\
&=&C_0\sum_{n=0}^\infty \left[ \sum_{m=0}^M(-\xi )^n\frac{M!S_{nm}^k(\beta )%
}{(M-m)!\sqrt{m!\Gamma (2k+m)/\Gamma (2k)}}\right] |n,k\rangle .  \nonumber
\end{eqnarray}
The combination of Eq.(34) and (39) gives the complete expansion of the LPS.

\section{Some su(1,1) optical systems}

In the previous two sections, we obtain the general results of several
quantum states for su(1,1) Lie algebra. Now we want to investigate some
interesting su(1,1) optical systems.

\subsection{Density-dependent HP realization}

The HP realization of the su(1,1) Lie algebra is \cite{HP}

\begin{equation}
K_{+}=a^{+}\sqrt{N+2k},K_{-}=\sqrt{N+2k}a,K_0=N+k.
\end{equation}
where $a^{+},a,$ and $N=a^{+}a$ are the creation, annihilation, and number
operator of a single-mode electromagnetic field satisfying $[a,a^{+}]=1.$ On
the Fock space $|n\rangle =[a^{+n}/\sqrt{n!}]|0\rangle ,$ we have

\begin{eqnarray}
K_{+}|n\rangle &=&\sqrt{(n+1)(2k+n)}|n+1\rangle , \\
K_{-}|n\rangle &=&\sqrt{n(2k+n-1)}|n-1\rangle ,  \nonumber \\
K_0|n\rangle &=&(n+k)|n\rangle .  \nonumber
\end{eqnarray}
In comparison with Eq.(2), we see that the HP realization gives rise to the
discrete representation of su(1,1) Lie algebra on the usual Fock space.
Therefore, by replacing the state $|n,k\rangle $ by $|n\rangle ,$ we recover
all the results in Sec.II and III.

By the replacement procedure described above, we obtain the PCS via HP
realization as

\begin{equation}
|\alpha ,M\rangle _{NB}=(1-|\alpha |^2)^{M/2}\sum_{n=0}^\infty {%
{M+n-1 \choose n}%
}^{1/2}\alpha ^n|n\rangle
\end{equation}
This is just the well-known negative binomial state(NBS)\cite{NBS}. Here $%
M=2k.$ Since the PCS admits displacement operator formalism, we naturally
obtain the displacement operator formalism of the NBS from Eq.(4)\cite{FUnbs}

\begin{equation}
|\alpha ,M\rangle _{NB}=\exp [\eta a^{+}\sqrt{N+2k}-\eta ^{*}\sqrt{N+2k}%
a]|0\rangle .
\end{equation}
The parameter $\eta $ is determined by the equation $\eta /|\eta |\tanh
|\eta |=\alpha .$

From Eq.(16), the ladder operator formalism of the NBS is written as

\begin{equation}
\frac 1{\sqrt{N+M}}a|\alpha ,M\rangle _{NB}=\alpha |\alpha ,M\rangle _{NB}
\end{equation}
As seen from the above equation, we conclude that the NBS is a NLCS in Fock
space as discussed in our previous paper\cite{WANGnbs}. In addition, the
su(1,1) displaced number states via HP realization are studied in detail by
Fu and Wang\cite{WangFu}.

It can be seen that some useful results of the NBS are conveniently
extracted from the general results for su(1,1) Lie algebra .

\subsection{Amplitude-squared realization}

The amplitude-squared su(1,1) is given by

\begin{equation}
K_{+}=\frac 12a^{+2},K_{-}=\frac 12a^2,K_0=\frac 12(N+\frac 12).
\end{equation}
The representation on the usual Fock space is completely reducible and
decomposes into a direct sum of the even Fock space ($S_0$) and odd Fock
space ($S_1$),

\begin{equation}
S_j=\text{span}\{||n\rangle _j\equiv |2n+j\rangle |n=0,1,2,...\},\text{ }%
j=0,1.
\end{equation}

Representations on $S_j$ can be written as\cite{LPS}

\begin{eqnarray}
K_{+}||n\rangle _j &=&\sqrt{(n+1)(n+j+1/2)}||n+1\rangle _j, \\
K_{-}||n\rangle _j &=&\sqrt{(n)(n+j-1/2)}||n-1\rangle _j,  \nonumber \\
K_0||n\rangle _j &=&(n+j/2+1/4)||n\rangle _j.  \nonumber
\end{eqnarray}
The Bargmann index $k=1/4$($3/4$) for even(odd) Fock space. From Eq.(4) we
see that the PCSs in even/odd Fock space are squeezed vacuum state and
squeezed first Fock state

\begin{eqnarray}
|\xi \rangle _{SV} &=&\exp (\frac \xi 2a^{+2}-\frac{\xi ^{*}}2a^2)|0\rangle ,
\\
|\xi \rangle _{SF} &=&\exp (\frac \xi 2a^{+2}-\frac{\xi ^{*}}2a^2)|1\rangle ,
\nonumber
\end{eqnarray}
respectively. The ladder operator formalisms of the squeezed vacuum state
and squeezed first Fock state are easily obtained from
Eq.(16)\cite{Siva98}

\begin{eqnarray}
\frac 1{N+1}a^2|\xi \rangle _{SV} &=&\xi /|\xi |\tanh (|\xi |)|\xi \rangle
_{SV,} \\
\frac 1{N+2}a^2|\xi \rangle _{SF} &=&\xi /|\xi |\tanh (|\xi |)|\xi \rangle
_{SF.}  \nonumber
\end{eqnarray}

We see that the the two states are the two-photon nonlinear coherent state $%
|\alpha \rangle _{TP}$ which is defined as

\begin{equation}
f(N)a^2|\alpha \rangle _{TP}=\alpha |\alpha \rangle _{TP}.
\end{equation}
Here $f(N)$ is a real function of the operator $N.$

Now we consider the matrix elements $S_{nm}^k(\xi )$ (Eq.(34)) in the
representation(Eq.(47)). Reminding that the Bargmann index $k=1/4(3/4)$ for
even(odd) Fock space and substituting $\alpha =\exp (i\theta )\tanh r$ into
the Eq.(34), we obtain the matrix elements in the representation as

\begin{eqnarray}
S_{nm}^{1/4}(\xi ) &=&\frac{(-1)^m}{m!n!}\sqrt{\frac{(2n)!(2m)!}{\cosh r}}%
\exp [i(n-m)\theta ](\tanh r/2)_{}^{m+n}\text{ }  \nonumber \\
&&_2F_1(-m,-n;1/2;-1/\sinh ^2r),
\end{eqnarray}

\begin{eqnarray}
S_{nm}^{3/4}(\xi ) &=&\frac{(-1)^m}{m!n!}\sqrt{\frac{(2n+1)!(2m+1)!}{\cosh
^3r}}\exp [i(n-m)\theta ](\tanh r/2)_{}^{m+n}  \nonumber \\
&&_2F_1(-m,-n;3/2;-1/\sinh ^2r).
\end{eqnarray}

As special cases of our general result(Eq.(34)), the above two equations
with $k=1/4(3/4)$ have been obtained by Marian\cite{Marian}.

\subsection{Two-mode realization}

The two-mode photon operators

\begin{equation}
K_{+}=a^{+}b^{+},K_{-}=ab,K_0=\frac 12(N_1+N_2+1)
\end{equation}
generate the su(1,1) Lie algebra. Here $N_1=a^{+}a$ and $N_2=b^{+}b.$ The
Fock space ${\cal F}$ of the two-mode states is decomposed into a direct sum
of irreducible invariant subspaces ${\cal F}_p^{\pm }$\cite{LPS}

\begin{eqnarray}
{\cal F} &=&{\cal F}_0\oplus {\cal F}_1^{\pm }\oplus ...\oplus {\cal F}%
_p^{\pm }\oplus ..., \\
{\cal F}_p^{+} &\equiv &\text{span}\{||n\rangle _{+p}\equiv |n,n+p\rangle
|n=0,1,2,...\},  \nonumber \\
{\cal F}_p^{-} &\equiv &\text{span}\{||n\rangle _{-p}\equiv |n+p,n\rangle
|n=0,1,2,...\}.  \nonumber
\end{eqnarray}
Representations on $F_p^{\pm }$ are isomorphic and take the form

\begin{eqnarray}
K_{+}||n\rangle _{\pm p} &=&\sqrt{(n+1)(n+p+1)}||n+1\rangle _{\pm p}, \\
K_{-}||n\rangle _{\pm p} &=&\sqrt{n(n+p)}||n-1\rangle _{\pm p},  \nonumber \\
K_0||n\rangle _{\pm p} &=&[n+(p+1)/2]||n\rangle _{\pm p}.  \nonumber
\end{eqnarray}
which are representation (2) with $k=(p+1)/2.$ Then by replacing $%
|0,k\rangle $ by $||0\rangle _{\pm p}$ and $k$ by $(p+1)/2$ in Eq.(4)$,$ we
obtain the two-mode squeezed vacuum state\cite{TWOSV}

\begin{equation}
|\xi ,p\rangle _{\pm }=\exp (\xi a^{+}b^{+}-\xi ^{*}ab)||0\rangle _{\pm p}
\end{equation}

From Eq.(16), ladder operator formalism of the two-mode squeezed vacuum
state is

\begin{equation}
\frac 2{(N_1+N_2)+p+2}ab|\xi ,p\rangle _{\pm }=\xi /|\xi |\tanh (|\xi |)|\xi
,p\rangle _{\pm }
\end{equation}
We can define two-mode NLCS as

\begin{equation}
f(N_1,N_2)ab|\alpha \rangle _{TM}=\alpha |\alpha \rangle _{TM}.
\end{equation}
Therefore, the two-mode squeezed vacuum state can be viewed as the two-mode
NLCS. In addtion, the pair coherent state\cite{Pair} is a special case of
two-mode NLCS with $f(N_1,N_2)=1.$

\section{Conclusions}

In this paper we have given the ladder operator formalism of a general
quantum state for su(1,1) Lie algebra. The algebra involved in the general
state is well-known GDO algebra. The ladder operator formalism of the PCS is
found and it is a su(1,1) NLCS. The expansion and exponential form of the
NLCS are given. The matrix elements of the su(1,1) squeezing operator is
obtained in terms of hypergeometric functions. Using the matrix elements,
expansions of the su(1,1) displaced number states and Laguerre polynomial
states are obtained. As realizations of su(1,1) Lie algebra, some
optical su(1,1) systems are considered. We obtain the ladder
operator formalism of the negative binomial state, squeezed vacuum state,
squeezed first Fock state, and two-mode squeezed vacuum state. We have
generalized the notion of the NLCS in Fock space to the su(1,1) case. It
is interesting to study further the su(1,1) NLCS in various quantum
optical systems.

\vspace{2cm}

{\bf Acknowledgment}: The author thanks for the discussions with Prof.
H. C. Fu and the help of Prof. C. P. Sun, S. H. Pan and G. Z. Yang. The
work is partially supported by the National Science Foundation of China
with grant number:19875008.


\begin{references}
\bibitem{CS}  R.J.Glauber, {\it Phys.Rev.Lett}. {\bf 10, }277(1963);\newline
R.J.Glauber, {\it Phys.Rev.} {\bf 130},2539(1963);\newline
R.J.Glauber, {\it Phys.Rev}. {\bf 131},2766(1963);\newline
W.M.Zhang, D.H.Feng and R.Gilmore, {\it Rev.Mod.Phys}. {\bf 62},867(1990).

\bibitem{MMCS}G. D$^{^{\prime }}$Ariano, M.Rasetti, and M.Vadacchino,
{\it Phys.Rev}. D{\bf 32},1034(1985);\newline
J.Katriel, M.Rasetti, and A.I.Solomon, {\it Phys.Rev}.D{\bf
35,}1284(1987);\newline
G.D$^{^{\prime }}$Ariano, S.Morosi, M.Rasetti, and
A.I.Solomon,
{\it Phys.Rev.}D{\bf 36},2399(1987);\newline
I.Jex and V.Buzek, {\it J.Mod.Opt}. {\bf 40},771(1993).

\bibitem{PCS}  A.Perelomov, {\it Generalized Coherent States and Their
Applications}(Springer-Verlag, Berlin, 1986).

\bibitem{BGCS}  A.O.Barut and L.Girardello, {\it Commun.Math.Phys}.{\bf \ 21,%
}41(1971).

\bibitem{MUS}  M.M.Nieto and L.M.Simmons,{\it \ Phys.Rev.Lett}. {\bf 41}%
,207(1978);\newline
M.M.Nieto and L.M.Simmons,{\it \ Phys.Rev.D}. {\bf 20},1321(1979);\newline
M.M.Nieto and L.M.Simmons, {\it Phys.Rev.D.} {\bf 20},1332(1979);\newline
M.M.Nieto and D.R.Truax, {\it Phys.Rev.Lett.} {\bf 71},2483(1993);

D. Trifonov, {\it J. Math. Phys}. {\bf 35}, 2297(1994).

\bibitem{LPS}  H.Y.Fan, X.Ye and Z.H.Xu, {\it Phys.Lett. A }199,{\bf 131}%
(1995).\newline
H.C.Fu and R.Sasaki, {\it Phys.Rev.A} {\bf 53},3836(1996).

\bibitem{HP}  T.Holstein and H.Primakoff, {\it Phys.Rev.} {\bf 58}%
,1098(1940).

\bibitem{ONE}  J.A.Bergou, M.Hillery and D.Yu, {\it Phys.Rev.A} {\bf 43}%
,515(1991).

\bibitem{TWO}  C.C.Gerry and R.Grobe, {\it Phys.Rev.A} {\bf 51},4123(1995).%
\newline
H.Y.Fan and X.Ye, {\it Phys.Lett.A} {\bf 175},387(1993).

\bibitem{GDO}  See, for example, H.Rampacher, H.Stumpf and F.Wagner {\it %
Fortschritte Phys.} {\bf 13},385(1965).

\bibitem{NLCS}  R.L.de Matos Filho and W.Vogel, {\it Phys.Rev.}A{\bf 54,}
4560(1996);

V.I.Man$^{,}$ko,G.Marmo,E.C.G.Sudarshan,\newline
and F.Zaccaria,{\it Physica Scripta}, {\bf 55} 528(1997);

O.V.Man$^{,}$ko,{\it Phys.Lett}.A{\bf 228},29(1997);

S.Mancini,{\it Phys.Lett}.A{\bf 233,}291(1997);

B.Roy,{\it Phys.Lett}.A{\bf 249},25(1998).

\bibitem{NBS}  K.Matsuo,{\it Phys.Rev.} A {\bf 41}, 519(1990);

A.Joshi and S.V.Lawande ,{\it J.Mod.Opt}{.} {\bf 38},2009(1991);

G.S.Agarwal, {\ }{\it Phys.Rev. A} {\bf 45}, 1787(1992);

S.M.Barnett ,{\it J.Mod.Opt.} {\bf 45}, 2201(1998);\newline
X.G.Wang and H.C.Fu, {\it Commun.Theor.Phys}. to appear.

\bibitem{FUnbs}  H.C.Fu and R.Sasaki ,{\it J.Phys.Soc.Japan} {\bf 66}%
,1989(1997).

\bibitem{WANGnbs}  X.G.Wang and H.C.Fu, {\it Mod.Phys.Lett.B}, {\bf 13},
 617(1999).

\bibitem{WangFu}  H.C.Fu and X.G.Wang, (unpublished).

\bibitem{Siva98} S. Sivakumar, {\it Phys. Lett. A} {\bf 250}, 257(1998).

\bibitem{Marian}  P.Marian, {\it Phys.Rev.A} {\bf 44}, 3325(1991);\newline
P.Marian, {\it Phys.Rev.A}{\bf \ 45}, 2044(1992).

\bibitem{TWOSV}  C.M.Caves and B.L.Schumaker,{\it \ Phys.Rev.A. }{\bf 31}%
,3068(1985).

\bibitem{Pair}  G.S.Agarwal. {\it J.Opt.Soc.Am}. {\bf B5}, 1940(1988);%
\newline
Ts.Gantsog and R.Tanas, {\it Opt.Commun}. {\bf 82},145(1991);\newline
S.-C Gou, J.Steinbach and P.L.Knight, {\it Phys.Rev.A} {\bf 54},4315(1996).
\end{references}
\end{document}